# How Pandemic Spread in News: Text Analysis Using Topic Model

WANG Minghao, Dr. Paolo MENGONI, IEEE Member

## Abstract

Researches about COVID-19 has increased largely, no matter in the biology field or the others. This research conducted a text analysis using LDA topic model. We firstly scraped totally 1127 articles and 5563 comments on SCMP covering COVID-19 from Jan 20 to May 19, then we trained the LDA model and tuned parameters based on the $C_V$ coherence as the model evaluation method. With the optimal model, dominant topics, representative documents of each topic and the inconsistence between articles and comments are analyzed. 3 possible improvements are discussed at last.

# Introduction

## Background

The World Health Organization (WHO) defined the SARS-CoV-2 virus outbreak as a severe global threat. The COVID-19 pandemic (hereinafter referred to as pandemic) has made tremendous impact on the whole world, including medical system, livelihood, employment and political issues. These issues have been projected to the media atmosphere significantly as well, which can be regarded as a media infodemic [1]. It is a precious chance to observe the topics that people are most interested in among the media during the pandemic using text analysis and topic modeling techniques.

## Related works

### *Topic model*

Topic model is a key technique to discover the latent semantic structure from text contents. This technique can be traced back to the latent semantic analysis (LSA) [2], which used singular-value decomposition to index and retrieve the semantic structure automatically. However, due to some basic statistical problems, a new approach called Latent Semantic Indexing (LSI) was proposed by Thomas Hofmann, which used statistical latent class model [3]. Latent Dirichlet Allocation was proposed then by Blei et al then as an extension of PLSA, which added Dirichlet priors for the document-specific topic mixtures [5] that has improved the model's generalization ability on unseen documents.

In the LDA, a *topic* was defined as a distribution over a fixed vocabulary [6]. For example, the *pandemic* topic has words about pandemic (e.g. *mask, vaccine*) with high probability, and the *politics* topic has words (e.g. *election, regime, lie*) about politics with high probability. These topics are determined before the document has been formed. The basic idea behind LDA was that documents usually bear multiple topics in different proportions. Each topic consists of multiple words in different proportions as well. The LDA model can capture documents' hidden structures (i.e. the topics, per-document topic distributions, per-document per-word topic assignments).

***Text analysis using topic model***

In the field of text analysis, more and more researches are using topic model to mine data from media contents [7]. Compared with traditional text analysis methods, topic model is effective on discover the latent message from especially short texts, which may convey rich meaning via self-defined hashtags or slangs.

**Research questions**

Inspired by previous research, we are curious about what are the media contents that most people are interested in during the pandemic. Rather than purely analyze user generated contents (UGC) on platforms like Twitter and Weibo, we are more interested in professional news websites (e.g. SCMP, BBC News, CNN) for mainly 2 reasons: first, long news articles offer richer corpus and that will help to train a more precise topic model and get more precise analysis results; second, comparing news articles together with comments may reveal the effect of information diffusion and agenda setting.

In this way, we should train and fine-tune an LDA model that suits the corpus the best.

**RQ1**: What is the optimal LDA model like based on the corpus of news articles and comments?

Specifically, we are curious that what contents appear the most frequently among the articles and comments? What are people exactly saying when they talk about the pandemic?

**RQ2**: What is the distribution of different topics among both the news articles and comments covering the pandemic?

Moreover, during surfing on these news websites, we noticed that some people are talking about B when the article is about A. An example is shown in Figure 1 and Figure 2. In this case, the article reported some recent situation about the pandemic, including where did the coronavirus come from, what is the riskiest way of contagion, and who are the most susceptible among the crowd. However, among the 4 comments (since there is one duplicate), 2 of them are accusing Wuhan for the pandemic and laughing at Chinese people, and 1 of them is suspecting the credibility of media diffusing pandemic messages. This inconsistence between news articles and comments can reflect the effects of information diffusion and the interaction between information sender and receiver.

**RQ3**: How inconsistent are the news articles and their comments? And what factors contribute to the inconsistence?

# Method

## Data collection

South China Morning Post (hereinafter referred to as SCMP) was selected as the data source of this research for mainly 3 reasons compared with other news websites accessible: first, it has comments posted, which takes a half of the data we need; second, the news on SCMP is inclusive and international enough, which will give us a relatively comprehensive global view of the pandemic information; third, SCMP is considered credible according to the research by the Centre for Communication and Public Opinion Survey, CUHK.

We chose the Python library BeautifulSoup to scrape articles and comments from SCMP's website. Some difficulties occurred. First, the index pages, which contained URLs of articles, were infinite rolling while no obvious page number existed, which made it difficult to parse the page in a static manner. However, by monitoring the XHR (XMLHttpRequest), there found a code grows (from 0 to 60, 20 a step) in the requested URL. Some other pages contained 13-digit timestamps as their indicator. So, with this pattern recognized, this problem was handled by requesting from those URLs with an apikey.

The second problem was that, at each article page, texts were hidden somewhere and parsing by tags didn't work. After observing the json data, texts were found hidden in "APOLLO_STATE" where sentences of the article were broken into fragments stored in dictionary-like blocks. Then there came the third problem that after parsing the json preliminarily, fragments from other articles blended together. The solution for these two problems was that, regular expression was used firstly to recognize those fragmented sentences, and an indicator for the end of an article was detected then. Finally, those sentences were glued together as an article.

The whole data set consisted of 1127 news articles and 5563 pieces of comments from Jan 20 to May 19, after dropping null values (e.g. unparseable pages). 6 attributes of the articles were collected, namely the news id, the title, the article text, the release time of the article, the collecting date and the URL it came from. 7 attributes of the comments were collected, namely the username, the raw comment, the cleaned comment, the date of that comment, the news id, the status that if this comment was a reply and the collecting date.

## Experiment

### *Data preprocessing*

Proper data preprocessing is of great significance to train a precise topic model. With the help of Python library nltk, we first tokenized the documents (i.e. both news articles and comments) by any space or punctuation. Then, all tokens regarded as stop words would be blocked out from the token list, which would be made into dictionary and corpus later. Besides all default stop words from nltk, 1042 new stop words (including integer numbers from 1 to 999) were added according to preliminary topic modeling experiments.

Stemmer was not utilized in this research due to mainly 2 reasons: first, after trying both porter stemmer and snowball stemmer in nltk, words would become hard to read after being stemmed (e.g. “china” is stemmed as “chin”, which is the same word means “jaw”), which would lead to difficulties and misunderstandings in interpreting topic modeling results; second, another stemmer lancaster seemed not available at this moment. No proper stemmer did we find so far.

After building preprocessed articles and comments into dictionary then corpus using Python library gensim, two data frames were formed for different purposes. The first one was following the order of original data, consisting of news id, document (both news articles and comments), release time, token and type (news articles are marked as type 0 and comments are marked as type 1). This data frame was made for future indexing and text analysis. The other one was shuffled in order to train topic model. Training set took 90% of the shuffled data because the amount of our data was relatively small and more data was needed for training. The other 10% was remained as test set.

***LDA model training, fine-tuning and evaluation***

There are 4 parameters in the gensim LDA model we would adjust, namely the number of topics, iteration, chunk size and passes. The number of topics is the most important parameter for a topic model, since too few topics will make the model imprecise (underfitting) while too many will lead to overfitting. The iterations refer to the maximum number to iterate through each document in the corpus to calculate the probability of each topic. Since the model is trained on one data chunk at a time, the chunk size refers to the number of documents in every training chunk. The passes refers to how many times the whole corpus is trained on. The model will be more precise with bigger iterations and passes, while it takes longer time to train the model as well. We would trade that off by choosing the minimum of them meeting enough precision.

Several commonly used evaluation methods were considered. As addressed in the original paper of LDA [4], perplexity was applied to evaluate the model performance. Perplexity reflects how uncertain that a word belongs to a specific topic category. The smaller the perplexity, the more precise is the model. However, some up-to-date experiments and research have revealed that this perplexity is confused to some extent.[8] Even Dr. Radim Řehůřek, the author of gensim, also thought there were some problems to measure LDA with perplexity. Coherence measures like NPMI or UMass have been widely deployed in many cases, while the $C_V$ coherence was proved to be the best performing one according the research by Michael Röder et al, which showed results of evaluating topic models the closest to human evaluation.[8] $C_V$ coherence is essentially an index measures the co-occurrence of the words extracted by the topic model. If those words from the same topic co-occurs very often (i.e. the $C_V$ coherence is high), the model is well performed.

In order to control variable, we assume these four parameters are decoupled, which means adjusting them separately will not affect other parameters. We will prove this assumption later.

To save training time, we firstly tried topic number from 2 to 50, 10 a step (iterations = 10, chunk size = 100, passes = 10). As shown in Figure 3, the $C_V$ coherence grows up mainly within 12 topics. Then, we tried topic number from 2 to 17, 1 a step (iterations = 10, chunk size = 100, passes = 10). As shown in Figure 4, the $C_V$ coherence grows up mainly within 7 topics, then it keeps relatively stable (there is a drop when topics = 8). The $C_V$ coherence of 7 topics

is 0.477.

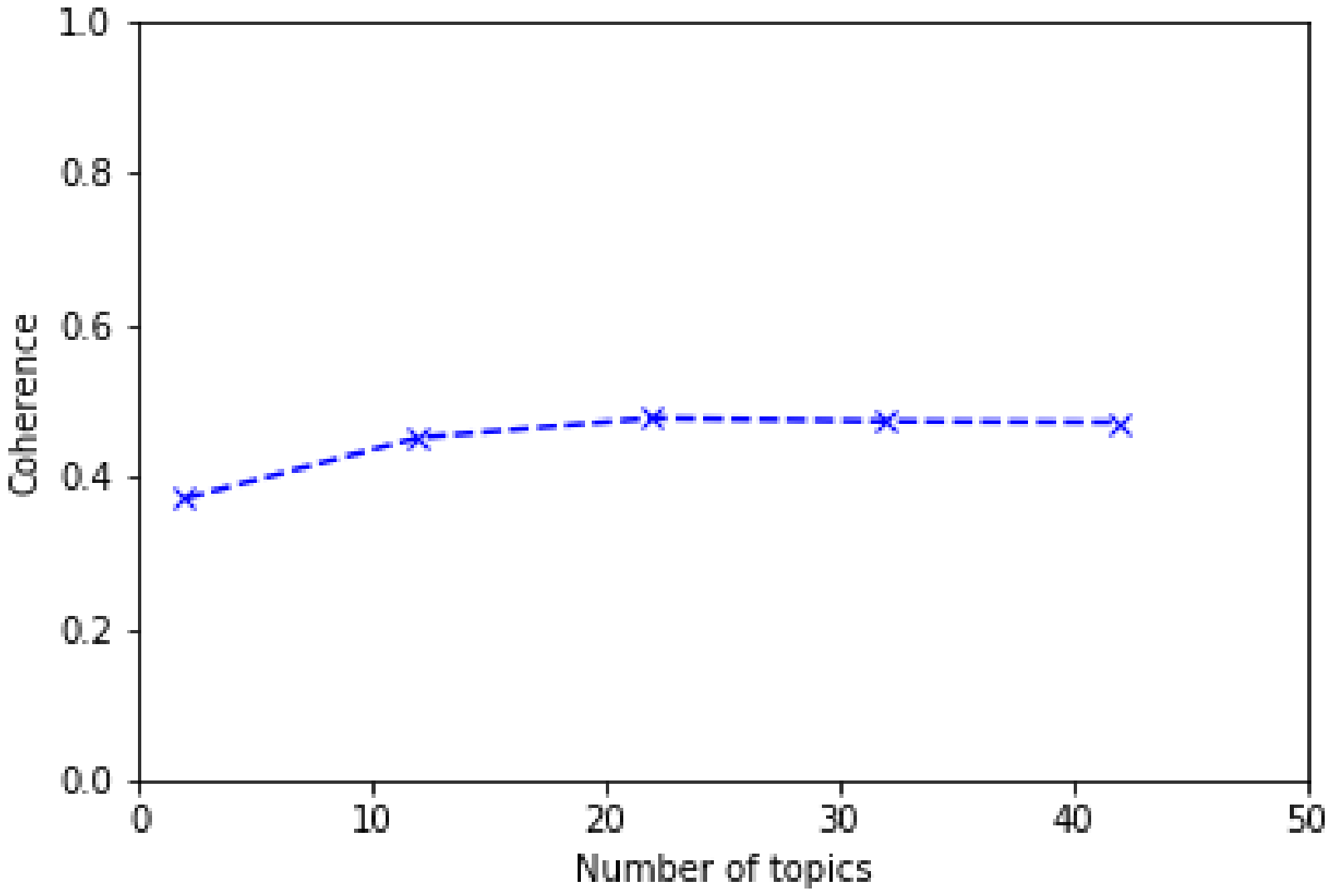

**Figure 3**

*The* $C_V$ *coherence on different number of topics (2 to 50, 10 a step).*

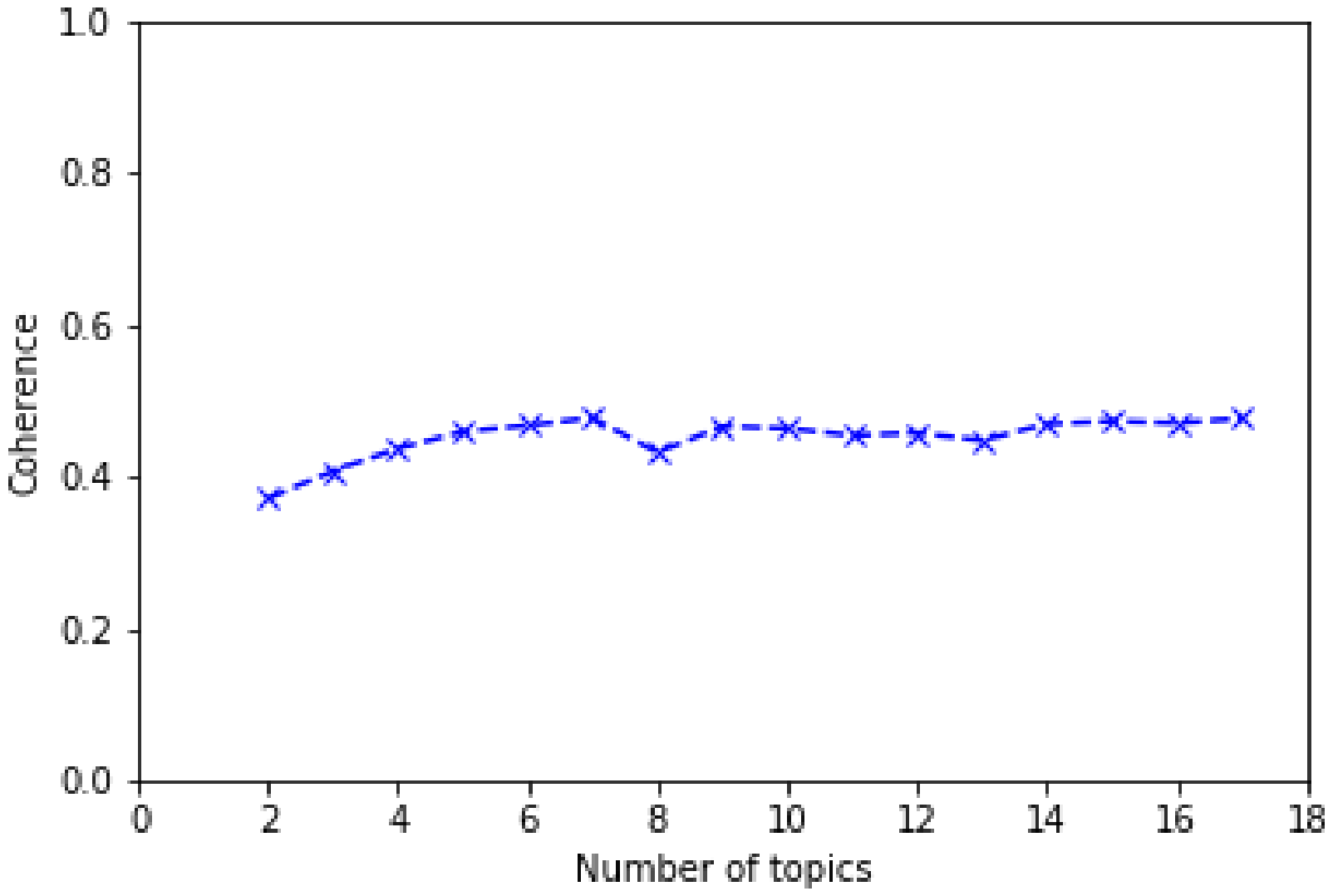

**Figure 4**

*The* $C_V$ *coherence on different number of topics (2 to 17, 1 a step).*

Different iterations were tested (i.e. 10, 100, 500, 1000) together with 7 topics. While the $C_V$ coherence didn't grow up as the iterations added. As for the chunk size and passes, no apparent grow as the parameters added. Results are shown in Table 1.

With these tuned parameters, we tested the model on the test set and got satisfactory results ($C_V$ coherence = 0.410).

About the decoupling assumption, we have also tested the same chunk size and passes with 12 topics, and then we calculated the Pearson correlation between the results, and they behaved strong correlations ( $r_{chunk\ size} = 0.857, r_{passes} = 0.735$ ), which proved those parameters can affect the model independently.

**Table 1**

*The $C_V$ coherence of different iterations, chunk size and passes.*

| parameter | parameter value | c_v coherence | other parameters |
|---|---|---|---|
| iterations | **10** | **0.439** | num_topics=7, chunksize=100, passes=10 |
| | 100 | 0.426 | |
| | 500 | 0.403 | |
| | 1000 | 0.366 | |
| chunksize | 10 | 0.381 | num_topics=7, iterations=100, passes=10 |
| | **100** | **0.460** | |
| | 500 | 0.422 | |
| | 1000 | 0.425 | |
| passes | **5** | **0.426** | num_topics=7, iterations=100, chunksize=100 |
| | 10 | 0.410 | |
| | 50 | 0.432 | |
| | 100 | 0.425 | |

***Topic analysis method***

With the well-tuned model, we firstly delivered a visualization work to give an overview of the whole data set using Python library pyLDAvis. Then, we calculated the dominant topic (i.e. the topic with the highest probability) of each document. We gathered those dominant topics and calculated the proportion of them. We also extracted the most representative documents of each topic by gathering the dominant topics separately and then picked up the document with the highest probability within that topic.

***Topic inconsistence evaluation method***

As addressed in the Introduction, we are going to compare the difference between articles and comments of the same news id, so an effective method to measure the inconsistence between them is needed. Considering every document had a list of topics (usually 7 digits) after applying that document with the optimal model, we delivered a small experiment on 3 measures (i.e. Spearman’s rank correlation coefficient, Kendall rank correlation coefficient and cosine similarity) to decide which measure to apply. We considered 2 pairs of topic lists, which is shown as Table 2. Each pair has only the first two digits different, and they should be considered very similar in the topic distribution. However, the results show that both Spearman’s and Kendall rank correlation coefficient are sensitive to the digits, while the cosine similarity is more stable that different digits share the similar ranks. In this way, cosine similarity is selected as the measure to evaluate the topic differences. The measure scheme is shown in Figure 5.

**Table 2**

*Comparison of different correlation measures on 2 pairs of similar topic lists.*

| | | |
|---|---|---|
| topic_list1 | [**1**, **2**, 0, **6**, 3, 4, 5] | Spearman_cor = 0.964, Kendall_tau = 0.905, **cosine_sim = 0.989** |
| topic_list2 | [**2**, **1**, 0, **6**, 3, 4, 5] | |
| topic_list3 | [**1**, **6**, 0, **2**, 3, 4, 5] | Spearman_cor = 0.107, Kendall_tau = 0.143, **cosine_sim = 0.725** |
| topic_list4 | [**6**, **1**, 0, **2**, 3, 4, 5] | |

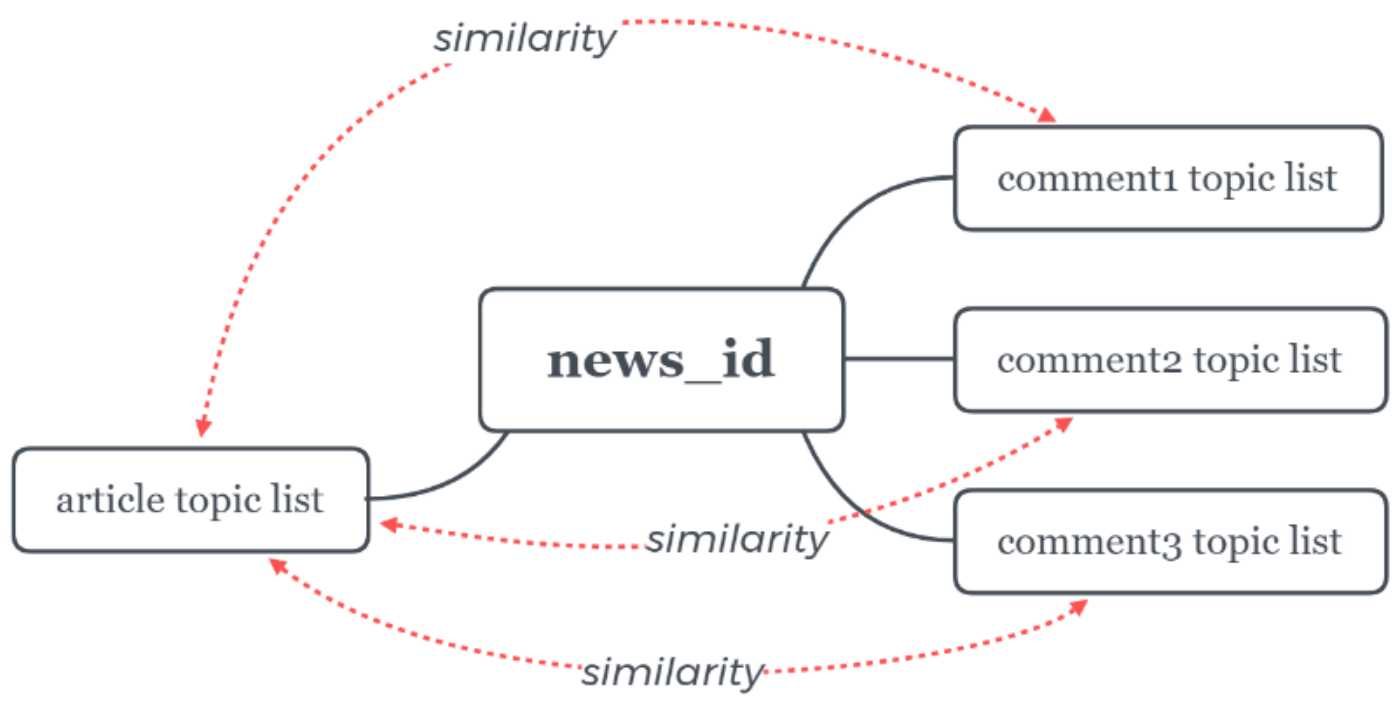

**Figure 5**

*A schematic diagram of the method to calculate the inconsistence between news articles and comments of the same news id.*

## Results

The optimal model has following parameters: number of topics = 7, iterations = 10, chunk size = 100, passes = 5.

### *Overview*

Table 3 shows the top 7 topic words within 7 topics. Words under each topic is described manually. As we can see from those topic words, the difference between topics are relatively obvious. Most of the words have actual meanings. It reflected the main agendas in SCMP covering COVID-19 from Jan to May. Interestingly, although this is an issue about health, while topics about health only take less than a half of all topics.

**Table 3**

*The top 7 words together with their probability within 7 topics. The descriptions are made manually.*

| topic number | top 7 topic words | | | | | | | description |
|---|---|---|---|---|---|---|---|---|
| 0 | 0.016*"patients" | 0.015*"covid" | 0.013*"disease" | 0.012*"wuhan" | 0.010*"vaccine" | 0.010*"coronavirus" | 0.009*"found" | Treatment |
| 1 | 0.011*"china" | 0.009*"$" | 0.007*"economy" | 0.007*"economic" | 0.006*"us" | 0.006*"companies" | 0.005*"million" | Commercial |
| 2 | 0.016*"workers" | 0.016*"hong" | 0.015*"kong" | 0.012*"home" | 0.010*"india" | 0.009*"singapore" | 0.008*"family" | International livelihood |
| 3 | 0.043*"us" | 0.015*"china" | 0.014*"taiwan" | 0.012*"war" | 0.010*"scmp" | 0.009*"comment" | 0.008*"military" | Conflict |
| 4 | 0.012*"cases" | 0.012*"new" | 0.010*"coronavirus" | 0.009*"000" | 0.008*"health" | 0.007*"city" | 0.007*"government" | Spread |
| 5 | 0.029*"china" | 0.026*"\" | 0.013*"world" | 0.009*"trump" | 0.008*"chinese" | 0.007*"even" | 0.006*"email" | Politics |
| 6 | 0.037*"china" | 0.023*"us" | 0.017*"chinese" | 0.011*"beijing" | 0.011*"pandemic" | 0.010*"world" | 0.009*"coronavirus" | China/US |

An overview graph of the topic distribution is shown in Figure 6. Notice that the numbers in the circles are different from the topic numbers. The overlapping of circles shows the distance (or closeness) between topics. The square of a circle refers to the total amount of that topic. We can find that circle 2, 3, 4, correspond with topic 1 (Commercial), 5 (Politics), 6 (China/US) respectively, are closely tied, which is consistent with our common sense. One word usually belongs to several topics. Some frequently applied words during the pandemic and their topic list (sorted by topic probability) are shown in Table 4.

**Table 4**

*Frequently applied words during the pandemic together with their belonged topics.*

| keyword | topic list (sorted) | | | | |
|---|---|---|---|---|---|
| coronavirus | 4 | 0 | 6 | 1 | |
| vaccine | 0 | 6 | | | |
| infections | 4 | 0 | | | |
| symptoms | 0 | 4 | | | |
| respiratory | 0 | | | | |
| government | 4 | 1 | 5 | 6 | 2 |
| lockdown | 4 | 2 | 1 | | |
| outbreak | 6 | 4 | 1 | 0 | |
| wuhan | 0 | 5 | 4 | 6 | |
| migrant | 2 | | | | |
| huawei | 3 | | | | |
| trade | 6 | 1 | 3 | | |

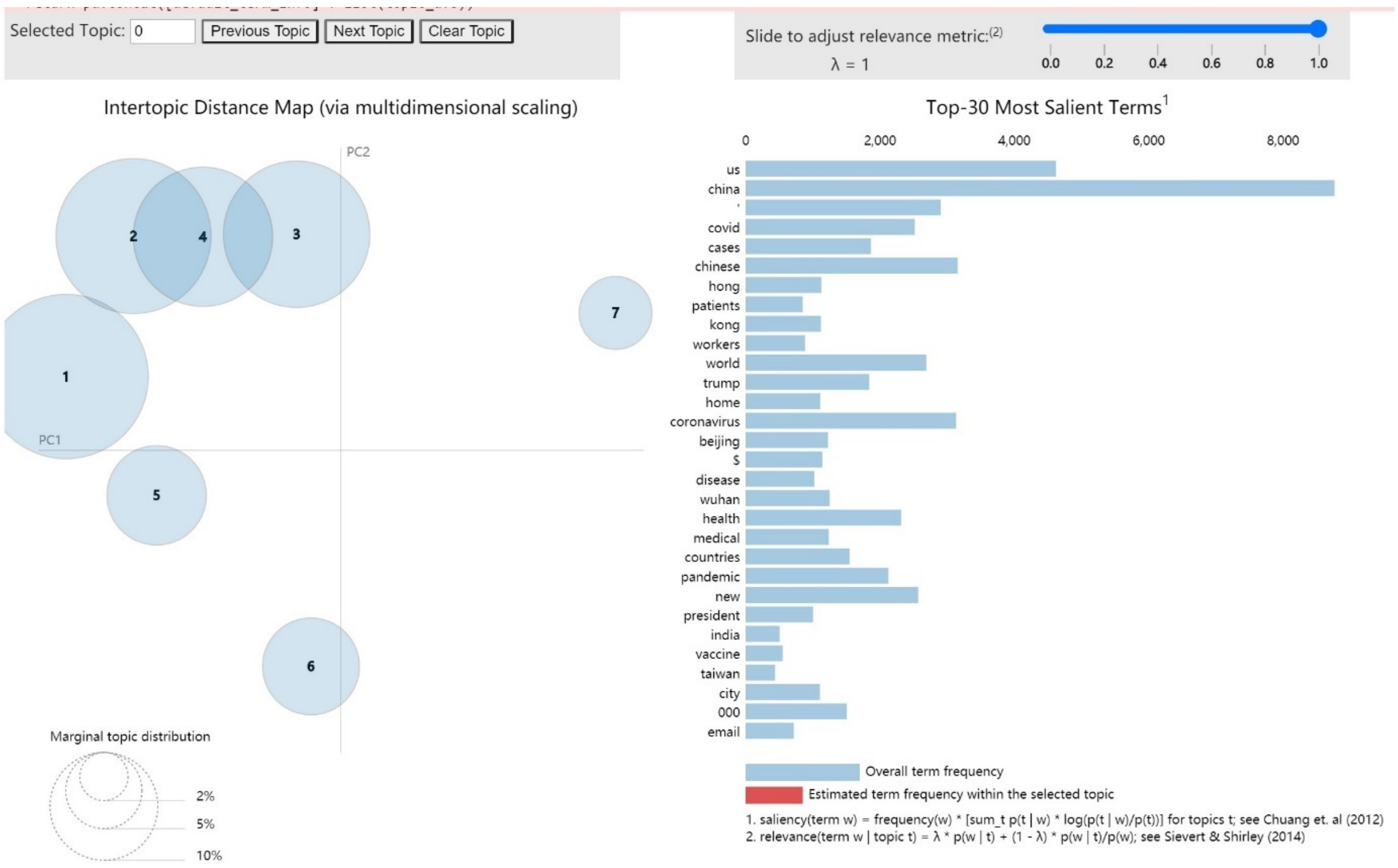

**Figure 6**

*An overview graph of the topic distribution.*

***Dominant topic analysis***

As shown in Figure 7, topic 5 (Politics) becomes the most popular of the pandemic report in SCMP from Jan to May, which takes more than a half of the agendas. Topic1 (Commercial) takes the second place among 7 topics with nearly 1/5 of the agendas. Topic 6 (China/US) takes nearly 10% of the agendas which is in the third place. Interestingly, articles and comments about treatment and virus itself merely take 0.5%. At the same time, very polarized conflict contents only appear 0.5% as well.

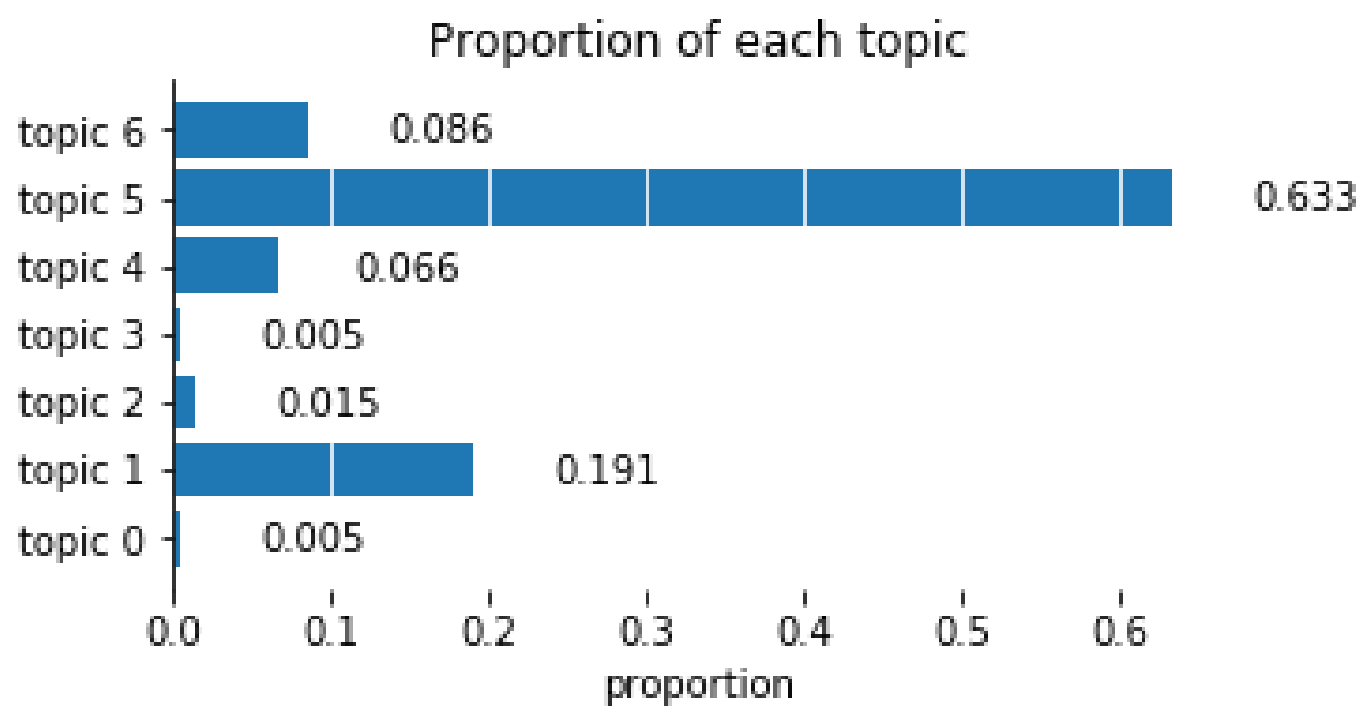

**Figure 7**

*The proportion of each topic as the dominant (i.e. the topic with the highest probability of a document).*

Table 5 shows the most representative contents of each topic. All the contents are happened to be comments.

**Table 5**

*The most representative contents of each topic. All the contents are happened to be comments.*

| |
|---|
| **TOPIC0 - 0.378, 2020-04-25** |
| *Guess their bank account is deep enough to cover the cost of their extended trip, as they seem not to be too concerned with a little pandemic-related hiccup in their RTW itinerary.* |
| **TOPIC1 - 0.573, 2020-05-03** |
| *@Werner Ziegler Actually no.... International scientists determined "Type A" were mostly found in the US and Australia. China is mainly Type B.* |
| **TOPIC2 – 0.523, 2020-04-29** |
| *The CCP is laughable. The rest of the world can see through its lies. and responsibility. avoidance immediately. What a joke the CCP is.* |
| **TOPIC3 – 0.379, 2020-04-20** |
| *how can you be cutting trips even primary sch children know that we are fighting dangerous contengious virus yet they cut trips.....?* |
| **TOPIC4 – 0.491, 2020-05-06** |
| *Dr. Fauci is the only credible person on that podium. Americans trust Fauci more than they trust Trump or Pompeo.* |
| **TOPIC5 – 0.701, 2020-05-18** |
| *"'They told us it wasn't contagious': Chinese blogger Fang Fang's forbidden diary reveals how Wuhan authorities told people coronavirus could NOT be passed between people" Fang FangSEARCH "Fang Fang Daily Mail"* |
| **TOPIC6 - 0.506, 2020-03-23** |
| *Trump is using China as a target to shift the public's attention because the government, under his leadership is failing the American people big time!* |

***Topic inconsistence analysis***

As shown in Figure 8, articles and comments shared a relatively good inconsistence in general. News ids with similarity lower than 60% only took 34.3%. Within the news ids of low similarity (i.e. of high inconsistence), the distribution of their dominant topics is shown in Figure 9. In order to examine if there was any specific topic contribute to the inconsistence more, we calculated Pearson correlation coefficient between this distribution and the dominant topic distribution among all documents. High correlation ($r = 0.957$) exists, which means no specific topic leads to the inconsistence. We selected several articles and their corresponding comments and tried to get some clues about the reason for inconsistence. Figure 10 and Figure 11 shows the article and comment with the lowest similarity (0.418). The article was talking about the “slump” occurred due to the pandemic, and trading volume and trend were discussed as well. While the comment was concerned with the home office fashion. Essentially, this concern was stemmed from the economic “slump”: besides the lockdown, many people work at home during this period to save office rental. In this case, the article and the comment were potentially related.

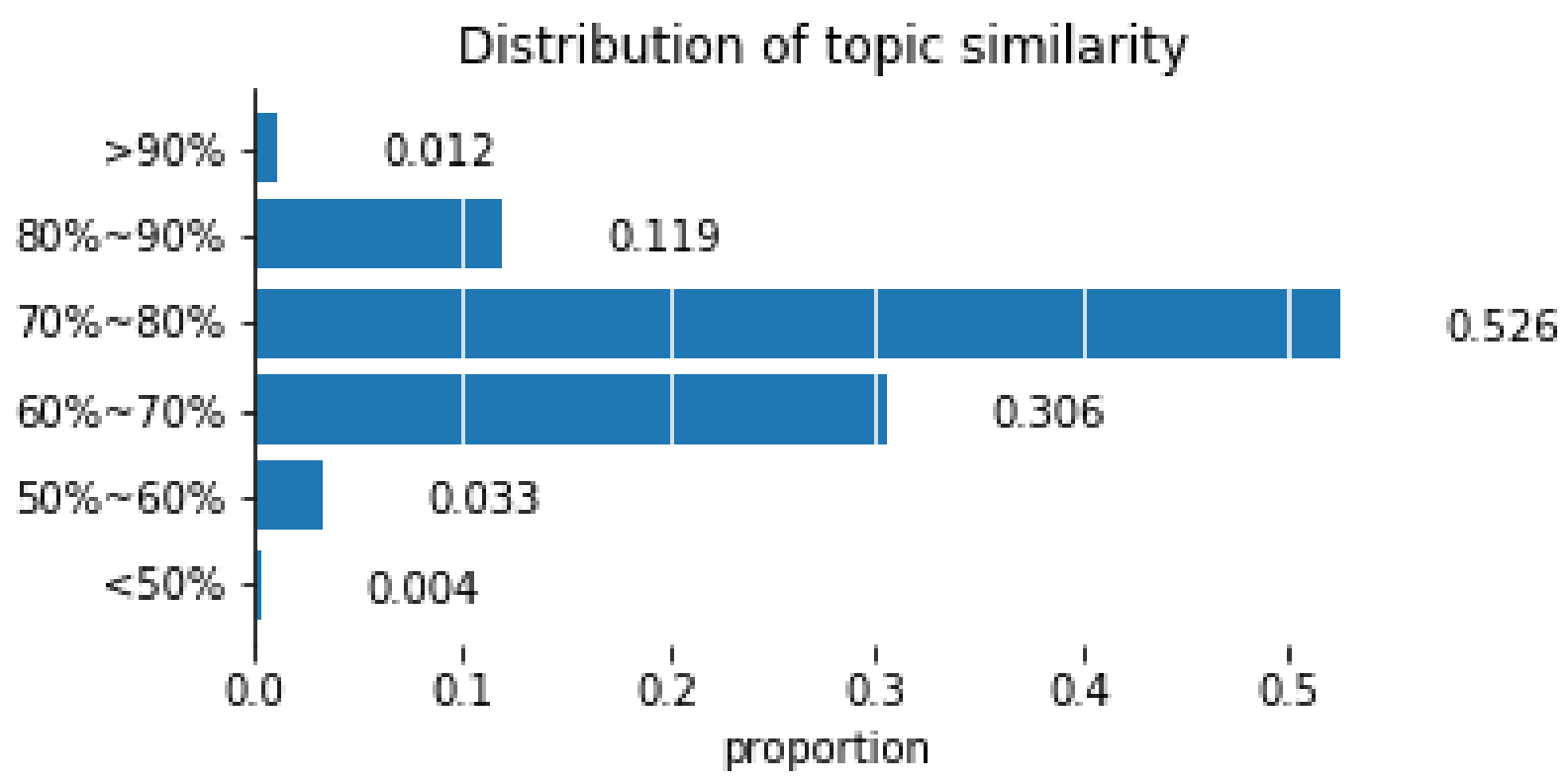

**Figure 8**
*The distribution of topic similarity within different ranges.*

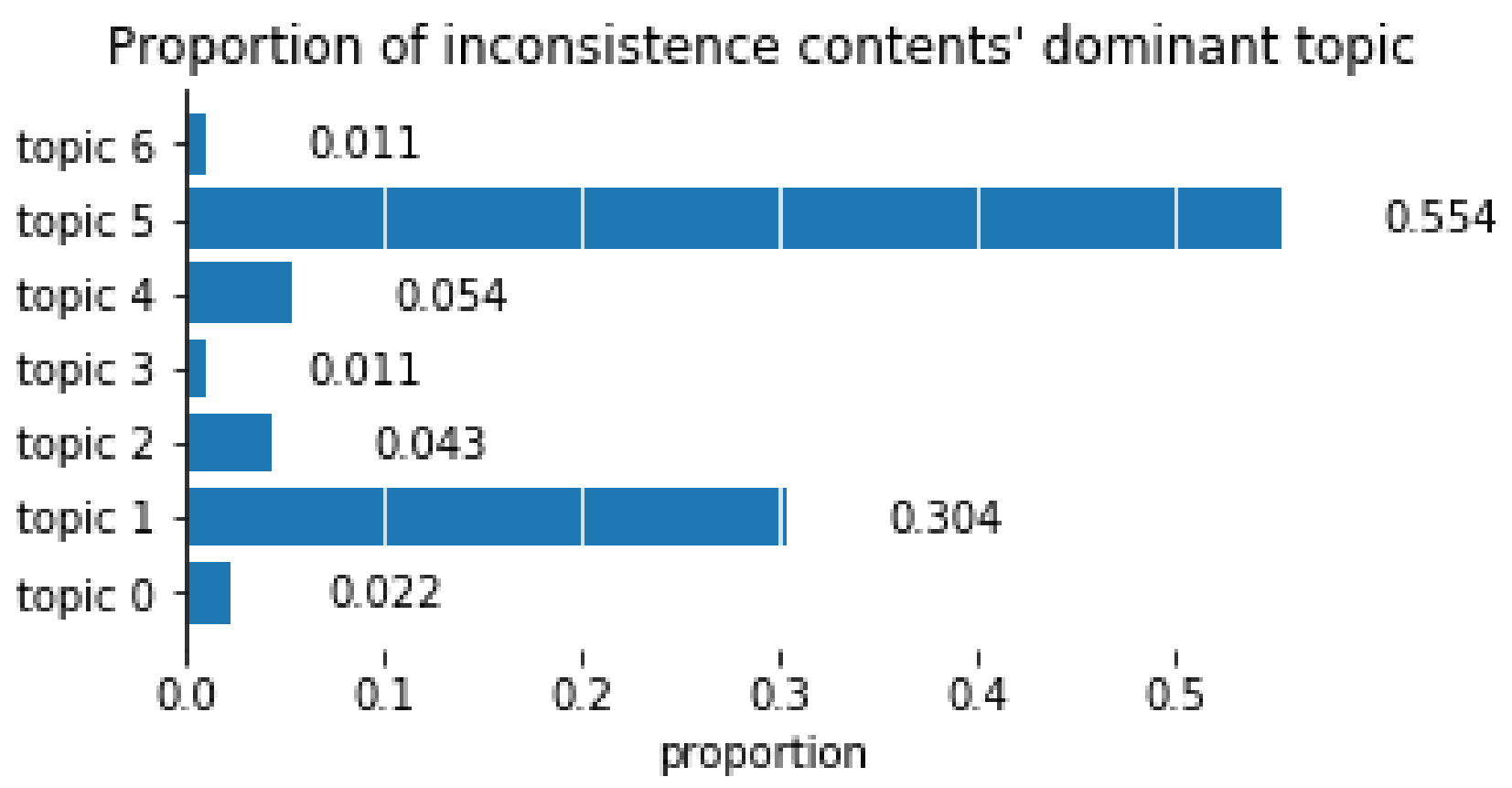

**Figure 9**

*The proportion of inconsistent documents grouped by their dominant topics.*

**3084240 - Article**↵

A slump in investment in Hong Kong and Singapore commercial property has contributed to the biggest pullback in a decade across Asia-Pacific. It may take another quarter before a hopeful rebound, JLL said. A slump in investment in Hong Kong and Singapore commercial property has contributed to the biggest pullback in a decade Investment fell by 50 per cent to US$21.3 billion last quarter from a year earlier, according to Real Capital Analytics, which tracks transactions worth at least US$10 million each. That is the least since the second quarter of 2010, it said. The drag reflects the "Many investors have paused activity due to the uncertain economic environment and, hence, deal activity has been impacted," said Stuart Crow, chief executive of capital markets in Asia-Pacific at JLL. "We see this reduced activity continuing into the second quarter, with trading volumes likely to bounce back more strongly in the second half of the year." Using a broader yardstick, JLL said investment in commercial properties in the Asia-Pacific region fell by more than a third to US$29.5 billion last quarter. A different measure used by CBRE showed a 25.4 per cent drop to near a three-year low of US$22 billion. Among six mature property markets in the region, Hong Kong fared the worst, with transaction volumes sliding by 74 per cent, according to JLL. They fell 68 per cent in Singapore and 61 per cent in mainland China. There were some stand-out transactions in the market, including Nonetheless, the first-quarter slump offers a warning to markets in Europe and the United States, whose economies were belatedly stricken by the pandemic. In the past week, governments from China to India have begun to reopen their economies as the outbreak comes under control. The slowdown in Hong Kong was exacerbated by the US-China trade war and the anti-government protests last year, according to JLL. Many investors have taken a wait-and-see approach, it added. Appetite for retail properties waned the most in Asia-Pacific last quarter, as transaction volumes fell by 39 per cent, the property consultants said. Office assets declined 35 per cent while hotel properties slipped 22 per cent. "The full impact of the Covid-19 outbreak on the investment market will hopefully start to become clearer in the second quarter, as investors focus on existing portfolios and bide their time for the right opportunities," said Regina Lim, executive director of capital markets research in Asia-Pacific at JLL. "Investors we speak to remain calm and optimistic."↵

**Figure 10**

*A case of topic inconsistence (article). Essentially, this comment is stemmed from this article.*

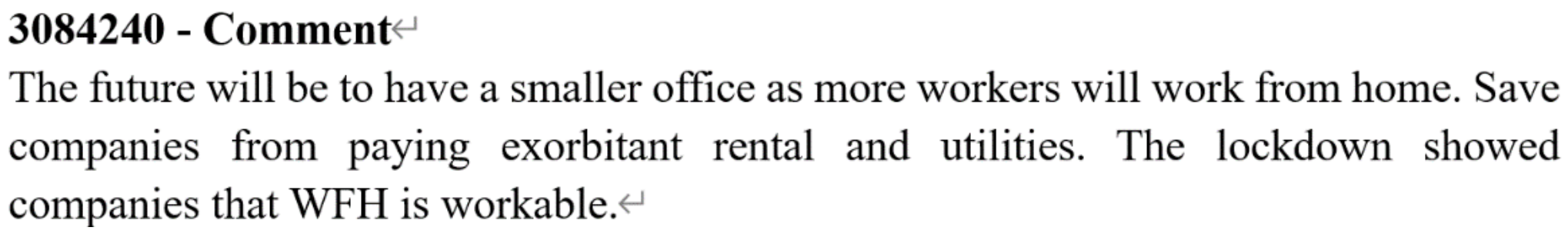

**3084240 - Comment**

The future will be to have a smaller office as more workers will work from home. Save companies from paying exorbitant rental and utilities. The lockdown showed companies that WFH is workable.

**Figure 11**

*A case of topic inconsistence (comment). Essentially, this comment is stemmed from this article.*

Figure 12 shows a relative obvious inconsistence case. This article was purely about the characteristics of the virus and some possible treatments. However, political accusation took a half part of the comments. Within all 4 comments, there was also one comment criticized media's spreading panic emotion. Only one comment was concerned with the article's words.

**3076022 – Article**

Researchers found that on average, infected people expel virus particles from their bodies for a relatively long period of 20 days, spreading it even before symptoms appear. A series of studies of the coronavirus suggest it is infectious for longer periods than pathogens from the same family, such as Sars, presenting added challenges for containing it. Researchers found that on average, people with SARS-CoV-2 – the virus that causes Covid-19, a pneumonia-like disease – can expel or “shed” virus particles from their bodies for a relatively long period of 20 days, spreading it even before symptoms appear. The virus also remains persistent in the faeces of some children, suggesting it can be transmitted through a faecal-oral transmission route – meaning that contaminated faeces from the infected child is somehow……

**3076022 – Comment1**

Thanks a lot, China, for infecting the world with the Wuhan Coronavirus. And thanks also for causing a global financial meltdown. The world will forever remember the recession of 2020 was caused by the Wuhan Coronavirus.

**3076022 – Comment2**

@Yum Cha 😂 just because your nick name is “Yum Cha” doesn’t make you ChineseAny more than a virus having skin colourFire up Spotify and listen to Zombie’s “ It’s in your head “ mate 😉

**3076022 – Comment3**

The funny thing about reading all media coverage is how bad they say this virus is but Infact it has killed less than SARS and less than even car accidents and most people recovered. 85% inflected gets a mild flu like symptoms like a fever and a sore throat and after a few days they recovered. Only those seriously sick are old and those with chronic disease anyway. Average age of the Italians that died are above 79 which is quite old and close to dying age anyway . I wish reporters can tell both sides of the story instead of adding to the negative and never reporting on the positive but I guess scared people read more news

**3076022 – Comment4**

In any area with questionable water quality (where what comes from the tap still requires boiling), following comparative discovery of negative tests from the front end of the human alimentary canal which indicate that COVID-19 can still remain active in fecal material (which comes out of the back-end of that same canal), the best advice is this: DO NOT skimp or short-change the time period that your potable water is boiled before you allow it to cool down or used, and keep that boiled water in a sealed container when not being tapped.

**Figure 12**

*Another case of topic inconsistence. Political issues are the most concerned with the comments, while the article itself is merely about medical issues.*

## Discussion

As we can see from the results, the articles and comments on SCMP covering COVID-19 from Jan to May are mainly concerned with topics like “politics” and “commercial” stemmed from the pandemic rather than the pandemic itself, and “politics” has attracted the most attention. Among topics related to the pandemic itself, people have paid much attention on the spreading situation while little on the treatment-related messages like “vaccine” or “ventilator”. We can describe these two features as people tend to “complain” and “panic”. Some people are passionate with accusing China as the “source of evil” and laughing at China’s telling lies.

Topic inconsistence between articles and comments existed to a normal extent, and no obvious topic was proved to lead the inconsistence. Some inconsistence was due to the very deep relationships between the article and comment, that means people may be associative in some cases. Meanwhile, we also found some inconsistence due to the emotional expression. When the news was talking about fact, people were often caring about political issues.

Our research still remains some work to be finished further in the future. For example, the SCMP provided news categorized by regions like “Greater China”, “North America” and “Europe”. These news articles have been already categorized by the editor, and different distribution among different regions could be discovered. Another one is about the data preprocessing. We were using unigram as our language model, and precision may be sacrificed for the speed. Some words like “Hong Kong” were broken down into “Hong” “Kong”, and this may lead to some kind of misunderstanding when generating the topics. Different from the unigram, bigram takes 2 words next to the keyword when calculating the conditional probability, thus it would be more precise to understand the contents. With enough computing power, bigram or even trigram should be tried in data preprocessing. Cosine similarity was applied in this research to measure the inconsistence between articles and comments. After some experiments, this method was the optimal among the options available now. But it is still sensitive to the topic number to some extent. In the future, a better measure remains to be discovered.

## Conclusion

Inspired by previous researches, this research applied LDA topic model to analyze the news articles and comments on SCMP covering the COVID-19 pandemic from Jan 20 to May 19. Model parameters were tuned on the training set after a series of experiments and tested on the test set. $C_V$ coherence, which was used as the measure of model performance, was good with the set of optimal parameters. Using the optimal model, the dominant topics from Jan to May was counted, the most representative contents were also analyzed, while several visualization works were conducted. Topic inconsistence phenomenon was measured with cosine similarity and the factors lead to this phenomenon were discussed. At last, 3 possible improvements were proposed.